\documentstyle[editedvolume,numreferences,epsf]{crckapb}

\def\be{\begin{equation}}
\def\ee{\end{equation}}
\def\bea{\begin{eqnarray}}
\def\eea{\end{eqnarray}}

\newcommand{\ket}[1]{|\kern.3ex#1\kern.3ex\rangle}
\newcommand{\bra}[1]{\langle\kern.3ex #1 \kern.3ex|}

\newcommand{\mean}[1]{\left\langle #1 \right\rangle} 
\newcommand{\smean}[1]{\langle #1 \rangle} 

\newcommand{\tr}[1]{\mathop{\mathrm{Tr}}\nolimits\left\{ #1 \right\}}  
  
\def\D{{\rm d}}                  

\begin{opening}

\title{Reversing the Sign of Current-Current 
Correlations}

\author{Markus  B\"uttiker}

\institute{D\'epartement de Physique Th\'eorique, Universit\'e de Gen\`eve,\\
CH-1211 Gen\`eve 4, Switzerland}

\end{opening}
\runningtitle{Current-Current Correlations} 
\begin{document}

\section{Introduction}

Dynamic fluctuation properties of mesoscopic electrical 
conductors provide additional information 
not obtainable through conductance measurement. 
Indeed, over the last decade, experimental and theoretical investigations
of current fluctuations have successfully developed into an 
important subfield of mesoscopic physics. A detailed report 
of this development is presented in the review by 
Blanter and B\"uttiker \cite{ybmb}. 

In this work we are concerned with the correlation 
of current fluctuations which can be measured
at different terminals of multiprobe conductors. 
Of particular interest are situations where, 
as a function of an externally controlled parameter, 
the sign of the correlation function can be reversed. 

Electrical correlations can be viewed 
as the Fermionic analog of the 
Bosonic intensity-intensity 
correlations measured in optical experiments.
In a famous astronomical experiment Hanbury Brown and Twiss
demonstrated that intensity-intensity 
correlations of the light of a star can 
be used to determine its diameter \cite{hanbury56}. 
In subsequent laboratory experiments of light 
split by a half-silvered mirror statistical 
properties of light were further 
analyzed \cite{hanbury57}. 
Much of modern optics 
derives its power from the analysis 
of correlations of 
entangled optical photon pairs 
generated by non-linear down conversion \cite{stein}. 
The intensity-intensity correlations of a thermal 
Bosonic source are positive due to statistical bunching. 
In contrast, anti-bunching of a Fermionic 
system leads to negative correlations \cite{pruc}.

Concern with current-current correlations in mesoscopic conductors 
originated with Refs. \cite{mb90,mb91}.  The aim of this work 
was to investigate the fluctuations and correlations 
for an arbitrary multiprobe conductor for which 
the conductance matrix can be expressed 
with the help of the scattering matrix \cite{mb86,mb88}. 
Refs. \cite{mb90,mb91} provided an extension of the discussions 
of shot noise by Khlus \cite{khlus}
and Lesovik \cite{lesovik} which applies to two-terminal conductors. 
These authors assumed from the outset that the transmission matrix
is diagonal and provided expressions for the two terminal shot 
noise in terms of transmission probabilities. 
It turns out that even for two probe conductors, 
shot noise can be expressed in terms of transmission 
probabilities only in a special basis (eigen channels). 
Such a special basis does not exist for multiprobe conductors
and we are necessarily left with expressions for shot noise 
in terms of quartic products of scattering matrices \cite{mb91,mb92a}.
There are exceptions to this rule: for instance correlations 
in three-terminal one-channel conductors can also be expressed 
in terms of transmission probabilities only \cite{tmrl}. 

The reason that shot noise, in contrast to conductance, 
is in general not simply determined by transmission 
probabilities is the following: if carriers incident from different 
reservoirs (contacts) or quantum channels can be scattered 
into the same final reservoir or quantum channel, quantum mechanics 
demands that we treat these particles as indistinguishable. We are 
not allowed to be able to distinguish from which initial 
contact or quantum channel a carrier has arrived. The 
noise expressions must be invariant under the exchange of
the initial channels \cite{mb92b,gram98,ybmb2,vlmb,esdl}.
The occurrence of exchange terms 
is what permitted Hanbury Brown and Twiss to measure 
the diameter of the stars: Light emitted by widely separated portions
of the star nevertheless exhibits (a second order) interference 
effect in intensity-intensity correlations \cite{hanbury56}. 

\begin{figure}[ht]
\vspace*{0.5cm}
\centerline{\epsfysize=5cm\epsfbox{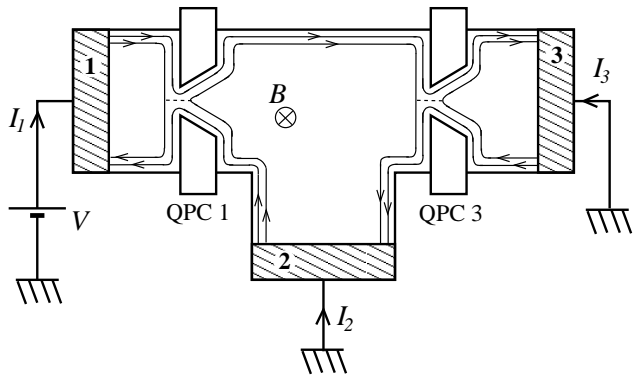}}
\vspace*{0.5cm}
\caption{\label{fig:exp}
Experimental arrangement of Oberholzer at al.
Current is injected at contact $1$. One edge channel
is perfectly transmitted (and noiseless) 
the other is partially transmitted at QPC~1 with probability $T_1$
and partially transmitted at QPC~3 with probability $T_3$ into contact 
$3$ and reflected with probability $R_3 = 1 -T_3$ into contact $2$. 
Of interest is the correlation of currents measured at contacts $2$ and $3$.}
\end{figure}

Experiments which investigate current-correlations 
in mesoscopic conductors have come along only recently. 
Oliver et al. used a geometry in which a "half-silvered 
mirror" is implemented with the help of a gate that
creates a partially transparent barrier \cite{oliver}. 
Henny et al. \cite{henny} separated 
transmission and reflection along edge states of a quantum point 
contact subject to a high magnetic field. In the zero 
temperature limit an electron reservoir compactly fills 
all the states incident on the conductor. 
A subsequent experiment
by Oberholzer at al. \cite{ober} uses a configuration with 
two quantum point contacts, as shown in Fig. {\ref{fig:exp}}. 
This geometry permits to thin out the occupation in the 
incident electron beam and thus allows to investigate 
the transition in the correlation 
as we pass from degenerate Fermi statistics to 
dilute Maxwell-Boltzmann satistics. Anti-bunching effects 
vanish in the Maxwell-Boltzmann limit and the current-current 
correlation tends to zero as the occupation of the incident 
beam is diminished. The fact that in electrical conductors 
the incident beam is highly degenerate is what made 
these Hanbury Brown Twiss experiments possible. In contrast, 
an emission of electrons into the vacuum generates an electron beam 
with only a feeble occupation of electrons \cite{kodama} and for this 
reason an experiment in vacuum has in fact just been achieved only 
very recently \cite{kiesel}. 
Below we will discuss the experiments in electrical conductors 
in more detail. 

Within the scattering approach, in the white noise limit, 
it can be demonstrated, 
that current-current correlations are negative, irrespective 
of the voltages applied to the conductor, temperature and 
geometry of the conductor \cite{mb91,mb92a}. The wide 
applicability of this statement might give the impression, 
that in systems of Fermions current correlations are always 
negative. However, the proof rests on a number of assumptions: 
in addition to the white-noise limit (low frequency limit) 
it is assumed that the terminals are all held at a constant 
(time-independent) terminal-specific potential. This is possible 
if the mesoscopic conductor is embedded in a zero-impedance 
external circuit. No general statement 
on the sign of correlations exists if the 
external circuit is characterized by an arbitrary impedance. 

In this work we are interested in situations for which the 
above mentioned proof does not apply. 
For instance, a voltmeter ideally has infinite impedance, 
and a conductor in which one of the contacts is connected to 
a voltmeter presents a simple example in which it is possible 
to measure {\it positive} current-current correlations \cite{ctmb}.  
In steady state transport the 
potential at a voltage probe floats to achieve 
zero net current. If the currents fluctuate the potential 
at the voltage probe must exhibit voltage fluctuations 
to maintain zero current at every instant. 
As has been shown by Texier and B\"uttiker, the fluctuating potential 
at a voltage probe can lead to a change in sign of 
a current-current correlation \cite{ctmb}.

A voltage probe also 
relaxes the energy of carriers, it is a source of 
dissipation \cite{mb85,mb86v,cbmb,liu}. 
Probes which are non-dissipative are of interest as models 
of dephasors. At low temperatures dephasing is quasi-elastic 
and it is therefore reasonable to model dephasing 
in an energy conserving way. This can be achieved
by asking that a fictitious voltage probe maintains 
zero current at every energy \cite{djbe}. 
Ref. \cite{vlmb} presents an application of this 
approach to noise-correlations in chaotic cavities. 

It is of 
interest to investigate current-correlations 
in the presence of such a dephasing voltage 
probe and to compare the result with a real dissipative 
voltage probe. 
No examples are known in which a 
dephasing probe leads to positive correlations. 
However, there exists also no proof that correlations 
in the presence of dephasing voltage probes are always negative. 

The proof that correlations in Fermionic conductors 
are negative also does not apply in the high-frequency 
regime. We discuss the frequency-dependence of equilibrium 
fluctuations in a ballistic wire to demonstrate 
the ocurrence of positive correlations at large 
frequencies. 

Another form of interactions which can induce positive 
correlations comes about if a normal conductor 
is coupled to a superconductor. 
Experiments have already probed shot noise in hybrid
normal-superconducting 
two-terminal structures \cite{jehl1,kozh,jehl2,reulet,lefl}. 
In the Bogoliubov
de Gennes approach the superconductor creates 
excitations in the normal conductor 
which consist of correlated electron-hole 
pairs. The process which creates the correlation 
is the Andreev reflection process by which an 
incident electron (hole) is reflected as a hole (electron). 
In this picture it is the occurrence 
of quasi-particles of different charge which makes 
positive correlations possible \cite{anda,thma,toma}. 
The quantum statistics remains Fermi like 
since the field operator associated with the Bogoliubov 
de Gennes equations obeys the commutation rules 
of a Fermi field \cite{gram99}. 
Alternatively the superconductor can be viewed as an injector 
of Cooper pairs \cite{rsl}. In this picture is the brake-up of Cooper pairs 
and the (nearly) simultaneous emission of the two electrons through 
different contacts which makes positive correlations possible. 
Our discussion centers on the conditions (geometries) 
which are necessary for the observation 
of positive correlations in mesoscopic normal conductors 
with channel mixing. Boerlin et al. \cite{boerlin} have investigated 
the current-correlations of a normal conductor 
with a channel mixing central island 
seprated by tunnel junctions from the contacts and the 
superconductor. 
Samuelsson and B\"uttiker \cite{psmb1} 
consider a chaotic dot which can have completely 
transparent contacts or contacts with tunnel junctions. 
Interestingly while a chaotic cavity with perfectly transmitting  
normal contacts and an even wider perfect contact to the superconductor 
exhibits positive correlations, application of a magnetic flux 
of the order of one flux quantum only is sufficient to destroy 
the proximity effect and is sufficient in this particular geometry 
to change the sign of correlations from positive to negative \cite{psmb1}. 
Equally interesting is the result that a barrier at the 
interface to the superconductor helps to drive the 
correlations positive \cite{psmb1}. 

\section{Quantum Statistics and the sign of Current-Current Correlations}

In this section we elucidate the connection between 
statistics and current-current correlations in 
multiterminal mesoscopic 
conductors and compare them with intensity-intensity 
correlations of a multiterminal wave guide 
connected to black body radiation sources \cite{mb91,mb92a}. 
We start by considering a conductor that is so small 
and at such a low temperature that transmission 
of carriers through the conductor can be 
treated as completely coherent. 
The conductor is embedded in a zero-impedance external 
circuit. Each contact, labeled $\alpha = 1, 2, ...$, is characterized 
by its Fermi distribution function 
$f_\alpha$.  Scattering of electrons at the 
conductor is described by a scattering matrix $S$. 
The $S$-matrix relates the incoming amplitudes to the outgoing amplitudes: 
the element $s_{\alpha\beta,mn}(E)$ gives the amplitude of the 
current probability in contact $\alpha$ in channel $m$ 
if a carrier is injected in 
contact $\beta$ in channel $n$ with amplitude 1 
(see \cite{mb92a} for a more precise 
definition). The modulus of an $S$-matrix element is the probability for 
transmission from one channel to another.
We introduce a {\it total} transmission probability 
(for $\alpha \ne \beta$)
\be
\label{trans}
T_{\alpha\beta} = 
\tr{s^{\dagger}_{\alpha\beta}(E) s_{\alpha\beta}(E)} 
\:.\ee
Here the trace is over transverse quantum channels and spin quantum numbers. 
This permits to write the conductance in the form \cite{mb86,mb92a} 
\be
\label{cond}
G_{\alpha\beta} = - \frac{e^2}{h} 
\int\D E\, (-df/dE) T_{\alpha\beta}
\:.\ee
where $f$ is the equilibrium Fermi function. 
The diagonal elements 
of the conductance matrix can be expressed with the help 
of $s_{\alpha\alpha}$. With the help of the {\it total}
reflection probability $R_{\alpha\alpha} = N_{\alpha} -
\tr{s^{\dagger}_{\alpha\alpha}(E) s_{\alpha\alpha}(E)}$
where $N_{\alpha}$ is the number of quantum channels 
in contact $\alpha$ 
we have $G_{\alpha\alpha} ={e^2}/{h} 
\int\D E\, (-df/dE)[N_{\alpha} - R_{\alpha\alpha}]$. 
Alternatively, since  $\sum_{\beta} G_{\alpha\beta} =
\sum_{\alpha} G_{\alpha\beta} = 0$ 
the diagonal elements can be obtained from the 
off-diagonal elements. 
The average currents of the conductor are determined 
by the transmission probabilities 
and the Fermi functions of the reservoir 
\be
\label{curf}
I_{\alpha} = \frac{e}{h} \int dE [(N_{\alpha} - R_{\alpha\alpha})f_{\alpha}
-\sum_{\alpha\beta} T_{\alpha\beta}(E) f_{\beta}]
\:.\ee 
In reality the currents fluctuate. The total current 
at a contact is thus the sum of an average current and a fluctuating 
current. We can express the total current in terms of a "Langevin" equation 
\be
\label{curff}
I_{\alpha} = \frac{e}{h} \int dE [(N_{\alpha} - R_{\alpha\alpha})f_{\alpha}
-\sum_{\alpha\beta} T_{\alpha\beta}(E) f_{\beta}] + \delta I_{\alpha} 
\:.\ee 
We have to find the auto - and cross-correlations of the fluctuating 
currents $\delta I_{\alpha}$ such that at equilibrium 
we have a Fluctuation-Dissipation theorem and such 
that in the case of transport the correct non-equilibrium 
(shot noise)  is described by the fluctuating currents. 
The first part of Eq. (\ref{curff}) represents the 
average current only in the case that the Fermi distributions 
are constant in time. This is the case if the conductor 
is part of a zero-impedance external circuit. If the external 
circuit has a finite impedance, the voltage at a contact 
fluctuates and consequently the distribution function 
of such a contact is also time-dependent. In this section we 
consider only the case of constant voltages in all the contacts. 

We compare the current fluctuations of the 
electrical conductor with the intensity fluctuations of a 
(multi-terminal)
structure for photons in which each terminal connects 
to a black body radiation source characterized by a Bose-Einstein 
distribution function $f_\alpha$. Like the electrical conductor 
the wave guide is similarly characterized by scattering matrices 
$s_{\alpha\beta}(E)$. 

The noise spectrum is defined as 
$P_{\alpha\beta}(\omega)2\pi\delta(\omega+\omega')=
\smean{\delta\hat I_\alpha(\omega)\delta\hat I_\beta(\omega')
+\delta\hat I_\beta(\omega')\delta\hat I_\alpha(\omega)}$
with $\delta\hat I_\alpha(\omega)=
\hat I_\alpha(\omega)-\smean{\hat I_\alpha(\omega)}$, where 
$\hat I_\alpha(\omega)$ is the Fourier transform of the current 
operator at contact $\alpha$. The zero frequency limit 
which will be of interest here is denoted by: 
$P_{\alpha\beta} \equiv P_{\alpha\beta}(\omega = 0)$.
The scattering approach leads to the following expression 
for the noise \cite{mb90,mb91,mb92a}
\be
\label{noise}
P_{\alpha\beta}=\frac{2e^2}{h}\int\D E\, 
\sum_{\gamma,\lambda}
\tr{A_{\gamma\lambda}^\alpha A_{\lambda\gamma}^\beta} 
f_\gamma (1 \mp f_\lambda)
\:.\ee
The matrix $A_{\lambda\gamma}^{\beta}$ is composed 
of the matrix elements of the 
current operator in lead $\beta$ associated 
with the scattering states describing carriers 
incident from contact $\lambda$ and $\gamma$ and is given by 
\be
\label{amatrix}
A_{\gamma\lambda}^\alpha=\delta_{\alpha\gamma}\delta_{\alpha\lambda}
-s^\dagger_{\alpha\gamma}(E)s_{\alpha\lambda}(E)
\:.\ee
In Eq. (\ref{noise}) the upper sign refers to Fermi statistics 
and the lower sign to Bose statistics. 

To clarify the role of statistics it is useful 
to split the noise spectrum in an equilibrium like part
$P^{eq}_{\alpha\beta}$ and a transport part 
$P^{tr}_{\alpha\beta}$ such that $P_{\alpha\beta} =
P^{eq}_{\alpha\beta} + P^{tr}_{\alpha\beta}$. 
We are interested in the correlations of the currents 
at two different terminals 
$\alpha \ne \beta$. 
The equilibrium part consists of Johnson-Nyquist  noise contributions 
which can be expressed in terms of transmission probabilities 
only \cite{mb91,mb92a} 
\be
\label{peq}
P^{eq}_{\alpha\beta} = 
- \frac{2e^2}{h}\int\D E\, (T_{\alpha\beta} f_\beta (1 \mp f_\beta)
+ T_{\beta \alpha} f_\alpha (1 \mp f_\alpha)
\:.\ee
Since both for Fermi statistics and Bose statistics 
$f_{\alpha}(1 \mp f_{\alpha}) = -kT df_{\alpha}/dE$
is positive, the equilibrium fluctuations 
are {\it negative} independent of statistics. 
The transport part of the noise correlation 
is 
\be
\label{ptr}
P^{tr}_{\alpha\beta }= \mp \frac{2e^2}{h}\int\D E\, \sum_{\gamma,\lambda}
\tr{s^{\dagger}_{\alpha\gamma}
s_{\alpha\lambda}
s^{\dagger}_{\beta\lambda}
s_{\beta\gamma}} f_\gamma f_\lambda
\:.\ee
To see that this expression is negative for Fermi statistics 
and positive for Bose statistics one notices that 
it can be brought onto he form \cite{mb91,mb92a} 
\be
\label{ptr1}
P^{tr}_{\alpha\beta }= \mp \frac{2e^2}{h}\int\D E\,
\tr{[ \sum_{\gamma} s_{\beta\gamma} s^{\dagger}_{\alpha\gamma} f_\gamma ]
[\sum_{\lambda} 
s_{\alpha\lambda}
s^{\dagger}_{\beta\lambda}f_\lambda ]} 
\:.\ee
The trace now contains the product of two self-adjoint matrices. 
Thus the transport part of the correlation has a definite 
sign depending on the statistics. 

It follows that current-current correlations 
in a normal conductor are negative due to the Fermi statistics 
of carriers whereas for a Bose system we have the possibility 
of observing positive correlations, as for instance 
in the optical Hanbury Brown Twiss experiments \cite{hanbury56,hanbury57}. 

There are several important assumptions which 
are used to derive this result: It is assumed that
the reservoirs are at a well defined chemical potential.
For an electrical conductor this assumption holds 
only if the external circuit has zero impedance.  
The above considerations are also valid only 
in the white-noise (or zero-frequency limit). 
We have furthermore assumed that the conductor
supports only one type of charge, electrons 
or holes, but not both. Below we are interested 
in examples in which one of these assumptions 
does not hold and which demonstrate that also 
in electrical purely normal conductors we can, under certain 
conditions, have positive correlations.

\section{Coherent Current-Current Correlation}

We now consider the specific conductor shown in Fig. {\ref{fig:exp}.
It is a schematic drawing of the conductor used in the 
experiment of Oberholzer et al. \cite{ober}. 
The sample is subject to a high magnetic field such 
that the only states which connect one contact to another one 
are edge states \cite{hal82,but88}. We consider first the case when there
is {\it only one edge state} (filling factor $\nu =1$
away from the quantum point contacts). The edge state is partially transmitted 
with probability $T_1$ at the left quantum point contact 
and is partially transmitted with probability $T_3$ at the 
right quantum point contact. The potential $\mu_{1} = \mu + eV$ at contact $1$ 
is elevated in comparison with the potentials $\mu_{2} = \mu_{3} =\mu$ 
at contact $2$ and $3$. Thus carriers  enter the conductor at 
contact $1$ and leave the conductor through contact 
$2$ and $3$. 
Application of the scattering 
approach requires also the specification of phases. 
However, for the example shown here, without closed paths, 
the result is independent of the phase accumulated 
during traversal of the sample and the result can be 
expressed in terms of transmission probabilities only.

At zero temperature 
we can directly apply Eq. (\ref{ptr1}) to 
find the cross-correlation. 
Taking into account that only the energy interval 
between $\mu_{1}$ and  $\mu$ is of interest 
we see immediately that 
$P_{23}= \mp \frac{2e^2}{h} |eV|
[ s_{31} s^{\dagger}_{21}
s_{21} s^{\dagger}_{31}] $ which is equal to 
$P_{23}= \mp \frac{2e^2}{h} |eV|
[ s^{\dagger}_{21}s_{21} s^{\dagger}_{31}s_{31}] $.
But  $s^{\dagger}_{21}s_{21} = T_1 R_3$, where $R_3 = 1-T_3$ and 
$s^{\dagger}_{31}s_{31} = T_1 T_3$ and thus 
\be\label{ObRes}
S_{23}= - \frac{2e^2}{h} |eV| T_1^2 R_3 T_3 
\:.\ee

Transmission through the first quantum point contact 
thins out the occupation in the transmitted edge state.
This edge state has now an effective distribution
$f_{eff} = T_1$. The correlation function 
has thus the form $S_{23}= - \frac{2e^2}{h} |eV| f_{eff}^2 R_3 T_3$.
For $T_1 =1$ we have a completely occupied beam of carriers incident 
on the second quantum point contact and the correlation 
is maximally negative with $S_{23}= - \frac{2e^2}{h} |eV| R_3 T_3$.
In this case the correlation is completely determined by current 
conservation: Denoting 
the current fluctuations at contact $\alpha$
by $\delta I_{\alpha}$ we have $\delta I_1 +\delta I_2 + \delta I_3 = 0$.
Consequently since the incident electron stream is noiseless
$\delta I_1 = 0$ we have $P_{23} = - P_{22} = - P_{33}$. 
Therefore if the first quantum point is open the 
weighted correlation $p_{23} = P_{23}/(P_{22}P_{33})^{1/2} = -1$.
The fact that an electron reservoir is noiseless is an important 
property of a source with Fermi-Dirac statistics \cite{henny}. 

If the transmission through the first quantum point contact 
is less than one the diminished occupation 
of the incident carrier beam reduces the correlation. 
Eventually in the non-degenerate limit $f_{eff}$ 
becomes negligibly small and the correlation 
between the transmitted and reflected current 
tends to zero. This is the limit of Maxwell-Boltzmann 
statistics. 

The experiment by Oberholzer et al. \cite{ober} 
measured the correlation for the entire range of 
occupation of the incident beam and thus illustrates 
the full transition from Fermi statistics to Maxwell-Boltzmann
statistics. The experiment by Oliver et al. \cite{oliver}
even though it is for a different geometry 
(and at zero magnetic field) is discussed by the authors 
in terms of the same formula Eq. (\ref{ObRes}). 
The range over which the contact which determines the filling of 
the incident carrier stream can be varied is, however, more limited 
than in the experiment by Oberholzer et al..

Before continuing we mention for completeness 
also the auto-correlations  
\be
P_{33}=\frac{2e^2}{h} |eV| T_3T_1(1-T_3T_1) 
\:,\ee
\be
P_{22}=\frac{2e^2}{h} |eV| T_1R_3(1-T_1R_3) 
\:,\ee
For $T_1 =1$ this is the partition noise of a quantum point 
contact \cite{reznik,kumar}.

We are now interested in the following question: 
Carriers along the upper edge of the conductor 
have to traverse a long distance from quantum point contact $1$
to quantum point contact $3$ (see Fig. \ref{fig:exp}). 
How would quasi-elastic scattering (dephasing) or inelastic 
scattering affect the cross correlation Eq. (\ref{ObRes})? 
For the case treated above
where only one edge state or a spin degenerate 
edge is involved the answer is simple: the cross correlation 
remains unaffected by either quasi-elastic or inelastic scattering. 
The question (asked by B. van Wees) becomes interesting 
if there are two or more edge states involved. 
It is for this reason that Fig. (\ref{fig:exp}) 
shows two edge channels. 

\begin{figure}[ht]
\vspace*{0.5cm}
\centerline{\epsfysize=5cm\epsfbox{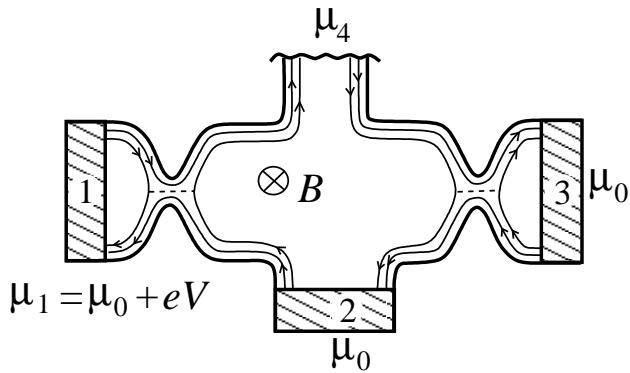}}
\vspace*{0.5cm}
\caption{\label{fig:dephase}
A voltage probe at the upper edge generates inelastic scattering 
or dephasing depending on whether the total instantaneous current 
or additionally the current at every energy is set to zero.  
After Texier and B\protect\"uttiker \protect\cite{ctmb}.}
\end{figure}

\section{Cross correlation in the presence of quasi-elastic scattering}

Incoherence can be introduced into the coherent scattering approach 
to electrical conduction with the help of fictitious 
voltage probes. (see Fig. \ref{fig:dephase}). Ideally a voltage probe 
maintains zero net current at every instant of time. 
[ A realistic voltmeter will have a finite response time. 
However since we are concerned with the low-frequency limit 
this is of no interest here.] A carrier entering a voltage probe 
will thus be replaced by a carrier entering the conductor from the voltage 
probe. Outgoing and incoming carriers are unrelated in phase and 
thus a voltage probe is a source of decoherence. 
A real voltage probe is dissipative. If we wish to model 
dephasing which at low temperatures is due to quasi-elastic scattering 
we have to invent a voltage probe which preserves energy. 
de Jong and Beenakker \cite{djbe} proposed 
that the probe keeps not only the total current zero 
but that the current in each energy interval is zero at every instant 
of time. Noise correlations in the presence of a dephasing voltage probe 
have been investigated by van Langen and the author for 
multi-terminal chaotic cavities \cite{vlmb}. 

In the discussion that 
follows we will assume, as shown in Fig. \ref{fig:exp}
that the outer edge channel is perfectly transmitted 
at both quantum point contacts. Only the inner edge channel 
is as above transmitted with probability $T_{1}$ at the 
first quantum point contact and with probability $T_{2}$
at the second quantum point contact. 
Elastic inter-edge channel scattering is very small
as demonstrated in experiments by van Wees et al. \cite{wees}, 
Komiyama et al. \cite{komiyama}, 
Alphenaar et al. \cite{alphenaar} and 
Mueller et al. \cite{mueller} and below we will 
not address its effect on the cross correlation. 
For a discussion of elastic interedge scattering in this 
geometry 
the reader is referred to the 
work by Texier and B\"uttiker \cite{ctmb}. We wish to focus
on the effects of quasi-elastic scattering and inelastic scattering.
The addition of the outer edge channel has no effect on the 
noise in a purely quantum coherent conductor. Edge channels 
with perfect transmission are noiseless \cite{mb90}. 

To model quasi-elastic scattering along the upper edge 
of the conductor we now introduce an additional 
contact (see Fig. \ref{fig:dephase}).  
To maintain the current at zero for each energy interval 
we re-write the Langevin equations 
Eq. (\ref{curff}) for each energy interval $dE$, 
\be
\label{elangevin}
\Delta I_{\alpha} (E, t) = 
\frac{e}{h} [(N_{\alpha} - R_{\alpha\alpha}(E))f_{\alpha}(E,t) 
-\sum_{\beta \ne \alpha} T_{\alpha\beta}(E) f_{\beta}(E,t)] + 
\delta I_{\alpha} (E,t) 
\:.\ee 
At the voltage probe we have $\Delta I_{4} (E, t) = 0$
and thus the distribution function of contact $4$
is given by 
\be
\label{dist}
f_{4}(E,t) = \bar f_{4}(E) + \delta f_{4} (E,t) 
\:,\ee
where the time-independent part of the distribution function 
is given by 
\be
\label{fphi}
\bar f_{4}(E) = 
\frac{1}{N_{\alpha} - R_{\alpha\alpha}} [\sum_{\alpha  = 1}^{\alpha = 3} 
T_{4\alpha}f_{\alpha}]  
\:,\ee 
and the fluctuating part of the distribution function is 
\be
\label{fphif}
\Delta f_{4}(E,t) = 
\frac{h}{N_{\alpha} - R_{\alpha\alpha}} \delta I_{4} (E,t) 
\:.
\ee 
Here we have taken into account that the distribution functions
at contact $1$, $2$ and $3$ are time-independent (equilibrium)
Fermi functions. Only the distribution at contact $4$ 
fluctuates. Additionally, its time-averaged part is 
a non-equilibrium distribution function. 
For the simple example considered here it is given by 
\be
\label{ad1}
\bar f_4 (E) = \frac{1+T_1}{2} f_1(E) + \frac{R_1}{2} f_2(E)
\:.\ee
It is a two step distribution function \cite{nagaev92,vlmb} 
as shown in Fig. {\ref{fig:dist}.
Since there is now also a fluctuating part of the distribution function 
the total fluctuating current 
at contact $\alpha$ contains according to Eq. (\ref{elangevin})
also a term $-(e/h) T_{\alpha\beta}(E) \Delta f_{4}(E,t)$. 
We take the transmission probabilities to be energy independent.
Integration over energy gives thus for the fluctuating current at contact 
$\alpha$ 
\be\label{Di}
\Delta I_\alpha= \delta I_\alpha - \frac{G_{\alpha4}}{G_{44}}\delta I_4
\:.\ee
As a consequence the correlation 
between the currents at contacts $\alpha$ and $\beta$ in
the presence of a quasi-elastic voltage probe is 
$P^{\rm qe}_{\alpha\beta}=\mean{\Delta I_\alpha\Delta I_\beta}$
with 
\be
\label{central} 
P^{\rm qe}_{\alpha\beta} = P_{\alpha\beta} - \frac{G_{\alpha4}}{G_{44}}
P_{\beta4} - \frac{G_{\beta4}}{G_{44}}P_{\alpha4} + 
\frac{G_{\alpha4}G_{\beta4}}{G_{44}^2}P_{44}
\:.\ee
Here $P_{\alpha\beta}$ are 
the auto-correlations and cross-correlations 
of the fluctuating currents in the energy resolved 
Langevin equation Eq. (\ref{elangevin}). 
The spectra are evaluated with the help of 
Eqs. (\ref{noise})
that apply for a completely coherent conductor 
except that we use the distribution functions 
$f_1, f_2, f_3$ and $\bar f_4 (E)$. 
In this procedure we neglect thus the fluctuations 
of the distribution function in the evaluation 
of the intrinsic noise powers $P_{\alpha\beta}$. 
[This is appropriate 
for the second order correlations of interest here, 
but not for the higher order cumulants \cite{nagaev,nagaev1}]. 

In contrast to Eq. (\ref{ptr1}) the current-current 
correlation Eq. (\ref{central}) 
is not necessarily negative. Taking into account that 
the off-diagonal conductances are negative and that the 
intrinsic spectra $P_{\alpha\beta}$ and $P_{4\beta}$
are negative for cross-correlations, it is clear that 
the first three terms in Eq. (\ref{central}) 
are negative. The forth term, due to the fluctuating distribution
in the dephasing contact, is positive. 
For all examples known to us, it turns out that 
for a dephasing voltage probe, the first three
terms win and the resulting correlation is 
negative \cite{vlmb}. 
For the inelastic (physical) voltage probe this is 
not the case as we demonstrate below. 

\begin{figure}[ht]
\vspace*{0.5cm}
\centerline{\epsfysize=4cm\epsfbox{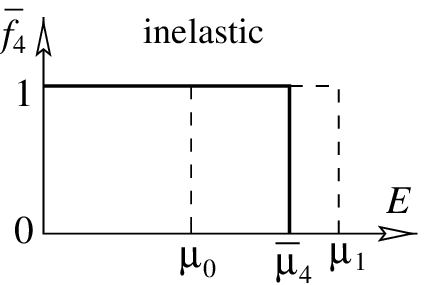}}
\hspace{0.25cm}
\centerline{\epsfysize=4cm\epsfbox{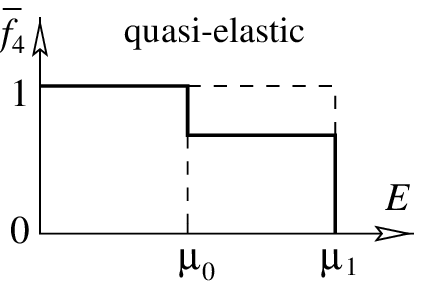}}
\vspace*{0.5cm}
\caption{
Distribution functions in the voltage probe reservoir.
For inelastic scattering this an equilibrium distribution 
function with a potential $\mu_4$ determined such that the 
average current vanishes. For a dephasing voltage probe 
this is a two step function determined such that the 
current at the probe vanishes at every energy. 
\label{fig:dist}}
\end{figure}

We will not discuss the most general result of Ref. \cite{ctmb} here 
but instead focus on the fact that such a dephasing probe 
can generate shot noise even in the case 
where the quantum coherent sample is noiseless. 

\section{Quasi-elastic partition noise} 

Consider the conductor for which the quantum point contacts 
are both closed for the inner edge channel $T_1 = 0$ and $T_3 = 0$. 
In this case, at zero temperature, the quantum coherent sample 
is noiseless. Transmission along each edge state is either one 
or zero. Now consider the conductor with the dephasing probe.
Under the biasing condition considered here, the distribution 
function $\bar f_4 (E)$ is still a {\it non-equilibrium} 
distribution function and given by 
$\bar f_4 (E) = \frac{1}{2} f_1(E) + \frac{1}{2} f_2(E)$.
The distribution 
function at the dephasing contact is similar to 
a distribution at an elevated temperature 
with $kT = eV/4$ with $eV$ the voltage applied between 
contact $1$ and contacts $2$ and $3$.
We have $\int dE \bar f_4 (E) (1 - \bar f_4 (E)) = e|V|/4$.
As a consequence the bare spectra  $P_{\alpha\beta}$ and 
$P_{\beta 4}$ are now non-vanishing.
Evaluation of the correlation function Eq. (\ref{central}) 
gives \cite{ctmb}, 
\be\label{Sqe}
P^{\rm qe}_{23}= -\frac{e^2}{h}|eV| \frac{1}{4}
\:.\ee
The electron current incident into the voltage probe 
from contact $1$ is noise-less. Similarly, the hole current 
that is in the same energy range incident from contact $2$ 
is noiseless. However, the voltage probe 
has two available out-going channels. The noise generated by 
the voltage probe is thus a consequence of the partitioning 
of incoming electrons and holes into the two out-going 
channels. In contrast, at zero-temperature, the  
partition noise in a coherent conductor is a purely 
quantum mechanical effect. Here the partioning
invokes {\it no quantum coherence} and is a 
classical effect. 

We emphasize that a dephasing voltage probe generates 
zero-temperature, 
incoherent partition noise whenever it is connected 
to channels which in a certain energy 
range are not completely filled. In our example the nonequilibrium 
filling of the channels incident on the voltage probe arises 
since the incident channels are occupied by 
reservoirs at different potentials. 
For instance a dephasing voltage probe connected 
to a ballistic wire (with adiabatic contacts) will generate 
incoherent partition noise if the probabilities 
for both left and right movers to enter the 
voltage probe are non-vanishing. If we demand that 
a dephasing voltage probe sees only left movers \cite{lith,emilio} 
(we might add a dephasing voltage probe 
which sees only right movers) we have a dephasing probe 
that not only conserves energy but also generates 
only forward scattering. As long as all incident channels 
are equally filled such a forward scattering 
dephasing probe will not generate partition noise. 

\section{Voltage probe with inelastic scattering} 

We next compare the results of the energy conserving voltage 
probe with that of a real (physical) voltage probe. 
At such a probe only the total current vanishes. 
The Langevin equations are 
\be
\label{flangevin}
I_{\alpha} (t) = \sum_{\beta} G_{\alpha\beta} V_{\beta}(t) 
+ \delta I_{\alpha} (t) 
\:.\ee 
where $G_{\alpha\beta}$ are the elements 
of the conductance matrix and $V_{\beta}(t)$ is the 
voltage at contact ${\beta}$. The voltages at contacts
$1, 2$ and $3$ are constant in time $eV_1 = \mu_1$ and 
$eV_{2} = eV_{3} = \mu_{0}$. But the voltage at contact 
$4$ is determined by $I_{4} (t) = 0$ and is given by 
 \be
\label{vtot}
V_{4}(t) = \bar V_{4} + \delta V_{4}(t) 
\:.\ee 
The time-independent voltage is 
\be
\label{vbar}
\bar V_{4} =  - G_{44}^{-1} \sum_{\beta \ne 4} G_{4 \beta} V_{\beta}
\:,\ee 
and the fluctuating voltage is 
\be
\label{vfluct}
V_{4}(t) =  - G_{44}^{-1} \delta I_{4} (t)
\:.\ee 
The distribution function in contact $4$ 
consists of a time-independent part 
and a fluctuating part. The time-independent
distribution is an {\it equilibrium} Fermi distribution 
at the potential $e\bar V_{4} = \bar \mu_4$. 
For our example we have 
\be
\label{vbarexp}
\bar V_{4} = \mu_{0} + \frac{1}{2} (1+T_{1}) e|V|
\:.\ee 
We remark that that this potential is independent of 
the transmission of quantum point contact 3.
The fluctuating currents at the contacts of the sample are 
\be
\label{ifluct}
\Delta I_{\alpha} (t) = \delta I_{\alpha} (t) - \frac{G_{\alpha 4}}{G_{44}}
\delta I_{4}
\:.\ee 
As a consequence the correlations of the currents 
are given by an equation which is similar to 
Eq. (\ref{central})
\be
\label{central1} 
P^{\rm in}_{\alpha\beta} = P_{\alpha\beta} - \frac{G_{\alpha4}}{G_{44}}
P_{\beta4} - \frac{G_{\beta4}}{G_{44}}P_{\alpha4} 
+\frac{G_{\alpha4}G_{\beta4}}{G_{44}^2}P_{44}, 
\:\ee
but with the important difference that the bare 
noise spectra are evaluated with the equilibrium 
Fermi functions $f_{\alpha}$ with $\alpha = 1, 2, 3, 4$. 

As in the quasi-elastic case, the first three terms are negative
and the forth term is positive due to the auto-correlations of 
the fluctuating voltage in this contact.  
In the inelastic case we can not consider the case 
where both $T_1$ and $T_3$ are zero since this implies 
that $\mu_4 = \mu_1$. Thus $T_1$ must be non-vanishing. 
On the other hand we are stil free to choose $T_3$ to simplify 
the problem. It is now interesting to consider the case $T_3 = 0$. 
In this case shot noise is generated at QPC $1$ and the voltage probe 
generates fluctuating populations in the the two out-going edge channels. 
The outer edge state leads carriers to contact $3$ and the inner edge 
state leads carriers to contact $2$. Interestingly, with this 
choice the first three terms in $P^{\rm in}_{23}$
vanish and the only non-zero term is the forth term 
arising from the auto-correlations of the 
voltage fluctuations in contact four. The correlation $P^{\rm in}_{23}$
is thus positive! 

In the presence of the voltage probe and for $T_3 = 0$, 
the correlation at contacts $2$ and $3$ is \cite{ctmb} 
\be
\label{posi} 
P^{\rm in}_{23} =  + \frac{e^2}{h}|eV| \frac{1}{2} T_{1} R_{1} 
\:.\ee
The autocorrelations are $P^{\rm in}_{22}  = P^{\rm in}_{33} =   
P^{\rm in}_{23}$. Current conservation 
is obeyed since $P^{\rm in}_{12} = P^{\rm in}_{13} = - 2P^{\rm in}_{23}$ and 
$P^{\rm in}_{11} = 4 P^{\rm in}_{23}$. 
Thus the 
normalized correlation function
is $p^{in}_{23} \equiv P^{\rm in}_{23} /(P^{\rm in}_{22} P^{\rm in}_{33})^{1/2}
= + 1$.  Clearly, as a consequence of the fluctuating voltage 
electrons are injected into the two edge channels 
in a correlated way. 

In the introduction we have remarked that thermal 
fluctuations are always anti-correlated. Therefore, 
as we increase the temperature in this conductor but 
keep the voltage fixed thermal fluctuations 
should eventually overpower the correlations due to 
the fluctating potential of the voltage probe. 
As a consequence, with increasing temperature, 
the correlation function should change sign. 
Indeed, a calculation gives \cite{ctmb} 
\bea
P^{\rm in}_{23}=\frac{e^2}{h}\bigg[ - k_{B}T
\left(2+{R_1}+{R_1T_1}\right) 
+\frac{R_1T_1}{2}{eV}\coth\frac{eV}{2k_{B}T}\bigg]
\:.\eea
If $k_{B}T=0$ we recover the positive result 
$P^{\rm in}_{23}= (e^2/h)|eV| R_{1} T_{1} /2$ for the shot noise, and
if $V=0$ we find 
$P^{\rm in}_{23}=- (2e^2/h)k_{B}T (1+ R_{1}/2)$, which is 
the result of the fluctuation-dissipation theorem:
$P^{\rm in}_{23}= 2 k_{B}T(G^{\rm in}_{23}+ G^{\rm in}_{32})$ 
where \cite{emilio}
$G^{\rm in}_{\alpha\beta}= G_{\alpha\beta}- (G_{\alpha4}G_{4\beta}/G_{44})$
is the conductance of the three-terminal conductor in the presence of 
incoherent scattering.

We define $T_c$, the critical temperature above which the correlations
$P^{\rm in}_{23}$ are negative. For small transmission $T_1\ll1$ we find:
$k_{B}T_c\simeq |eV| T_{1}/6$, and for large transmission $R_1\ll1$:
$k_{B}T_c\simeq |eV| R_{1}/4$. The transmission that maximizes the 
critical temperature is \cite{ctmb} 
$T_1=3-\sqrt6\simeq0.55$. In this case we have:
$k_{B}T_c^{\rm max}\simeq
|eV| (5\sqrt 6-12 )/(2(6\sqrt6-12))\simeq |eV|/21.8$.
Clearly, it would be intersting to see an experiment 
which investigates the reverasl of the sign of such a correlation function 
as a function of temperature. Another possibility is to perform 
the experiment at a fixed temperature but to make transmission into 
the voltage probe variable (for instance with the help of a gate). 
At temperatures so low that intrinsic inelastic scattering can be neglected 
the theory predicts a negative correlation if the connection is 
closed. As the contact to the voltage is opend there must exist 
a critical transmission probability at which the correlation 
vanishes. Finally for sufficiently large transmission 
the correlation is positive.

\begin{figure}[ht]
\vspace*{0.5cm}
\centerline{\epsfysize=5cm\epsfbox{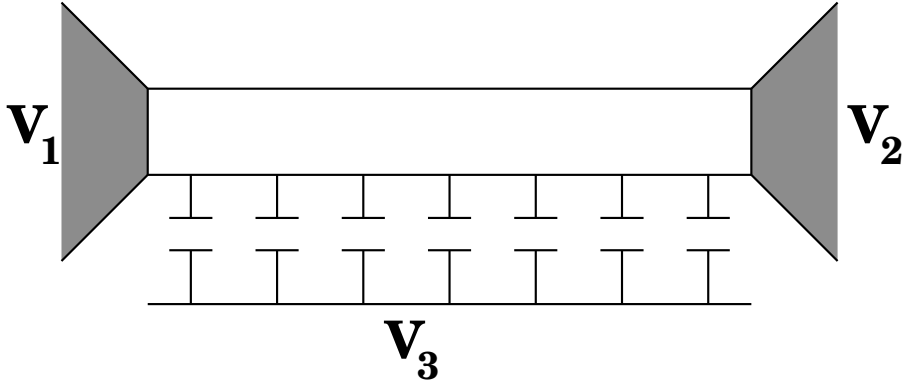}}
\vspace*{0.5cm}
\caption{\label{fig:blanter}
Ballistic one-channel wire coupled to reservoirs and capacitively 
coupled to a gate. After Blanter, Hekking and B\protect\"uttiker, 
\protect\cite{bhb}.
}
\end{figure}


\section{Dynamic Reversal of the sign of a Current-Current Correlation} 

The proof that current-current correlations in normal conductors 
are negative applies only to the white-noise (low-frequency)
limit. At finite frequencies it is possible to have positive 
current-current correlations even at equilibrium. 
To illustrate this we consider a one-channel 
ballistic conductor connected adiabatically to two 
electron reservoirs and capacitively coupled to a gate
with capacitance $c$ per unit length. The gate 
is connected to ground without additional impedance. 
The conductance matrix of this wire was calculated by 
Blanter, Hekking and B\"uttiker \cite{bhb} 
combining scattering theory with dynamic screening 
to determine the electrostatic potential self-consistently 
in random phase approximation. The conductance matrix 
determined in this way agrees with a theory based on 
a Tomonaga-Luttinger Hamiltonian and bosonization \cite{safi}.
The wire has a length $L$ and a density of states (per unit 
length) of $\nu_F = 2/hv_F$ where $v_F$ is the Fermi velocity. 
The interaction is described by the parameter 
\be
\label{g} 
g^{2} = \frac{1}{1+e^{2}\nu_F/c}
\:\ee
which is $1$ in the limit of a very large capacitance $c$
(non-interacting limit) and tends to zero as the 
capacitance $c$ becomes very small. 
The parameter $g$ and the density of states 
determine the static, electro-chemical capacitance \cite{mb93} 
of the wire vis-a-vis the gate, $c_{\mu} = g^{2} e^{2} \nu_F$. 
The dynamic conductance matrix is defined as 
$G_{\alpha\beta} (\omega) = 
\delta I_{\alpha} (\omega) /\delta V_{\beta} (\omega)$
where $\delta I_{\alpha} (\omega)$ and $\delta V_{\beta} (\omega)$
are the Fourier coefficients of the current at contact 
$\alpha$ and the voltage at contact $\beta$. Here $\alpha$
and $\beta$ label the contacts, the reservoirs $1$ and $2$ and 
the gate $3$. At equilibrium the dynamic conductance 
matrix is related to the current-current fluctuations 
via the Fluctuation-Dissipation theorem, 
\be
\label{fdt} 
P_{\alpha\beta}(\omega) = {\hbar \omega} \,\,coth{\frac{\hbar \omega}{2kT}}
G^{\prime}_{\alpha\beta} (\omega)
\:\ee
where $G^{\prime}_{\alpha\beta}(\omega) = 
(1/2) (G_{\alpha\beta}(\omega) + G^{\star}_{\beta\alpha}(\omega))$
is the real part of the element $G_{\alpha\beta}(\omega)$ of the 
conductance matrix. 
Consider now the current-current correlation $P_{12}(\omega)$. 
In terms of the effective wave-vector $q \equiv \omega g /v_F$ 
Ref. \cite{bhb} finds 
\be
\label{goff} 
G^{\prime}_{12} (\omega)
= - \frac{e^{2}}{h} \frac{16 g^{2} cos(qL)}{16 g^{2} cos^{2}(qL)
+ 4 (1 + g^{2})^{2} sin^{2}(qL)}
\:.\ee
In the zero-frequency limit we have 
$G^{\prime}_{12} (\omega) = - \frac{e^{2}}{h}$
which is negative as it must be for a conductance 
determined by a transmission probability. 
At a critical frequency 
\be
\label{omegac} 
\omega_{c} = v_F \pi/Lg  
\:.\ee
determined by $qL = \pi$, 
the real part of this conductance element becomes positive.
Hence for $\omega > \omega_c$ there 
exist frequency windows for which the equilibrium currents are positively 
correlated.  We note that in the non-interacting limit $g=1$ 
this frequency is determined by the time an electron 
takes to traverse half of the length of the wire, 
$\tau = L/2v_F$.  At this frequency the wire is charged 
by carriers coming in simultaneously from both reservoirs. 
Increasing the interaction suppresses charging and thus 
increases this frequency. On the other hand the frequency 
is inversely proportional to the length of the wire and the 
frequency tends to zero as the wire length tends to infinity. 

Since much of the discussion based on Luttinger theory 
and bosonization does not take into account the finite 
size of the sample, we can expect that such theories 
would in fact predict positive correlations! Indeed 
if we consider for a moment a Luttinger liquid 
coupled at a point $x = 0$ to a tunneling contact 
an electron inserted into the wire gives with probability $1/2$ rise 
to a left (right) going plasma excitation with charge e(1-g)/2 
and a right (left) going excitation with charge eg/2. This 
charges lead to positively correlated currents \cite{note2} at $x = \pm L/2$
with a noise spectrum proportional to $(1/4)g(1-g)$. 
Since the transition from a Luttinger liquid to a normal 
region leads to reflection of plasma excitations \cite{safi} 
we can expect that a proper treatment of the 
contacts would restore the expected negative 
correlations.

The positive dynamic correlations discussed here
are only accessible at high frequencies. Is it possible
to observe positively correlated currents at low frequencies? 
The answer is yes and we will now discuss two geometries.

\section{Positive Correlations of Dynamic Screening Currents}

Consider the classical electrical circuit 
shown in Fig. \ref{fig:xx} in which 
a node with potential $U$ is at one branch connected 
via a resistor with resistance $R$ to terminal $1$ 
and at the other branches via capacitances $C_{1}$ 
and $C_{2}$ to terminals $2$ and $3$. We are interested 
in the low-frequency behavior and expand the classical 
ac-conductance matrix ${\bf G}$ in powers of the frequency
\be
\label{class} 
{\bf G} (\omega) = -i \omega {\bf C} + \omega^{2} {\bf K} + ...
\:.\ee
The first term is purely capacitive with a capacitance 
matrix ${\bf C}$. The second term, which is of interest 
here, is the lowest order in frequency term which is dissipative. 
For the classical circuit of Fig. \ref{fig:xx} it is given by
\be
\label{Kmat}
{\bf K} = R
\left(
\begin{array}{ccc}
 C_{\Sigma }^{2}  &- C_{1}C_{\Sigma }& - C_{2}C_{\Sigma } \\
- C_{1}C_{\Sigma }& C_{1} ^{2}& C_{1}C_{2}\\
- C_{2}C_{\Sigma }&C_{1}C_{2}&C_{2} ^{2}
\end{array}
\right) \;\; .
\ee
Here $C_{\Sigma} = C_{1} + C_{2}$. 
The key point is of course that the off-diagonal elements 
$K_{23} = K_{32}$ are positive. In view of the fluctuation-dissipation 
theorem this implies that the correlation of currents 
at the two capacitive terminals are positively correlated, 
$P_{23} = 2 kT \omega^{2} K_{23} = 2 kT \omega^{2}C_{1}C_{2}R$. 
A current fluctation 
leads to charging of the capacitors $C_{1}$ and $C_{2}$ 
which in turn generates simultaneously current flowing through 
the two terminals $2$ and $3$ to compensate this charge. 

\begin{figure}[ht]
\vspace*{0.5cm}
\centerline{\epsfysize=4.5cm\epsfbox{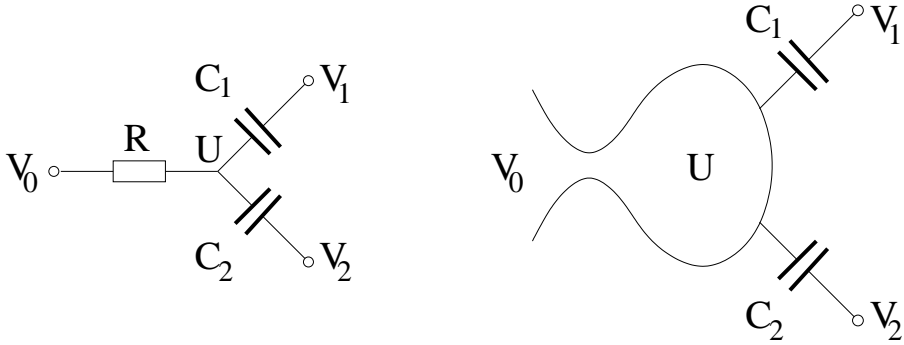}}
\vspace*{0.5cm}
\caption{\label{fig:xx}
Left: A classical three-terminal circuit: One branch is coupled 
via a resistor $R$ and two branches are coupled with capacitances
$C_1$ and $C_2$ to the node. Right: A cavity coupled via a narrow 
lead to a reservoir at voltage $V_{0}$ and coupled capacitively to two 
gates with geometric capacitances $C_1$ and $C_2$. $U$ is the voltage at 
node (left) and inside the cavity (right). 
}
\end{figure}

The classical circuit can, for example,  be viewed as a simple model 
for the ac currents of a mesoscopic (chaotic) cavity 
coupled capacitively to two gates with geometrical capacitances
$C_{1}$ and $C_{2}$ and connected via a quantum point contact 
to a particle reservoir. There are mesoscopic corrections 
to the geometrical capacitances and they are replaced by 
electrochemical capacitances $C_{\mu,1}$ and $C_{\mu,2}$. 
Similarly, the classical two terminal resistance $R$
is replaced by a {\it charge relaxation} resistance \cite{btp}. 
In a theory that determines the internal potential 
$U$ of the cavity in random phase approximation 
both the electrochemical capacitances and 
the charge relaxation resistances can be 
expressed in terms of elements of the 
Wigner-Smith, Jauch-Marchand, time delay matrix \cite{wigner,jauch,note}, 
\be
\label{Density Elements}
N_{\beta\gamma} = \frac{1}{2\pi i} \sum_{\alpha}
s_{\beta\alpha}^{\dagger} \frac{ds_{\gamma\alpha}}{dE}.
\ee
that characterizes fully the low-frequency charge
fluctuations on the cavity \cite{pedersen1}.
For the mesoscopic cavity the four dynamical transport 
coefficients of interest are, 
\be
\label{Four Parameters}
\begin{array}{cc}
D = e^2 \mbox{Tr} N, & C_{\mu,1} = \frac{C_{1} D}{ C_{\Sigma}+ D},\\ 
\quad\\
C_{\mu,2} = \frac{C_{2} D}{ C_{\Sigma}+ D}, &
R_q = \frac{h}{2e^{2}}
\frac{\left(\mbox{Tr} N^2\right)}{\left(\mbox{Tr} N\right)^2}.\\
\end{array}
\ee
Replacing $C_{1},C_{2}$ and $R$ in Eqs. (\ref{class}) and (\ref{Kmat}) 
with $C_{\mu ,1}, C_{\mu, 2}$ and $R_q$ gives 
the low frequency response of the mesoscopic cavity. 
Thus the equilibrium current correlations $P_{12} = P_{21}$
of the mesoscopic cavity are also positively correlated.

\begin{figure}[ht]
\vspace*{0.5cm}
\centerline{\epsfysize=5cm\epsfbox{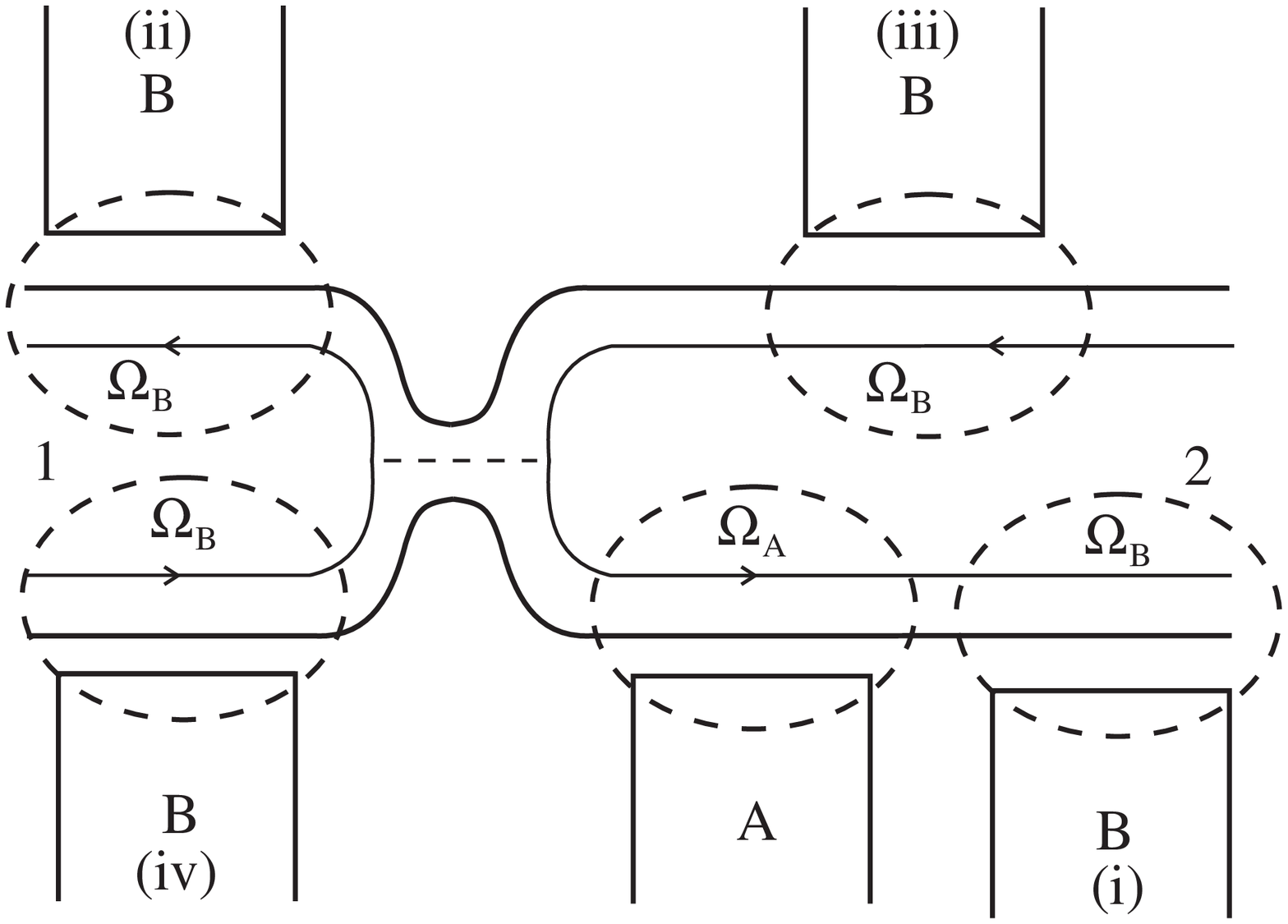}}
\vspace*{0.5cm}
\caption{\label{fig:ammb}
Quantum point contact in the quantum Hall regime: a single 
edge is partially transmitted and reflected. Charging of the edge 
state is probed capacitively with a capacitor at $A$ and 
second capacitor $B$ which can be in positions I-IV.
After Martin and B\protect\"uttiker \protect\cite{ammb}.}
\end{figure}
Thus far we have considered 
frequency dependent equilibrium fluctuations. 
Martin and B\"uttiker \cite{ammb} have investigated 
the correlation of dynamic screening currents 
in open conductors. 
The example considered is shown in Fig. \ref{fig:ammb}. 
A Hall bar in a high magnetic field is connected to two 
reservoirs $1$ and $2$.  The magnetic field corresponds 
to a filling factor $\nu =1$ such that the wire is in the 
integer quantum Hall regime. Backscattering is generated by 
a quantum point contact. The lines along the edges of the conductor 
indicate the edge states. Their chirality is indicated 
by arrows. 
Two gates $A$ and $B$ are used to 
probe charge fluctuations capacitively. Four positions $I, II, III, IV$
are considered for gate $B$ whereas gate $A$ is always held at the 
same position.  Charge fluctuations on an edge state will induce
capacitive currents on the gates $A$ and $B$. 

To keep the discussion simple it is assumed that charge pile up 
occurs only in the proximity of the gates $A$ and $B$ 
and that the remaining part of the conductor is charge neutral. 
The regions where charge pile up can occur are indicated by
the volumes $\Omega_A$ for gate $A$ and $\Omega_B$ for gate $B$. 
The geometrical capacitance of the gates to the edge states are 
denoted by $C_A$ and $C_B$. Again it is possible to 
express the charge fluctuations with the help 
of a generalized Wigner-Smith matrix. Whereas in the example 
of the cavity discussed above, the charge in the 
entire cavity was of interest, here we are interested only 
in the charge pile up in the regions $\Omega_A$ and $\Omega_B$. 
We are thus interested in {\it local} charge fluctuations. 
As a consequence we now have to consider functional 
derivatives of the scattering matrix with regard to the 
local potential \cite{mb96}, 
\begin{equation}
{\cal N}^{(\eta)}_{\delta \gamma} = \frac{-1}{2 \pi}
\sum_{\nu} \int_{\Omega_{\eta}} d^3 {\bf r} 
\left[ s^{\star}_{\nu \delta}(E, U({\bf r}))
\frac{\delta s_{\nu \gamma}(E, U({\bf r}))}{e\delta U({\bf
r})} \right] 
\label{d0}
\end{equation}
where $\eta = A$ or $\eta =B$ and 
${\bf r}$ is in the volume $\Omega_{\eta}$ and $U({\bf
r})$ is the electrostatic potential at position ${\bf r}$. For
example ${\cal N}^{(A)}_{1 2}({\bf r})$ is the electron density,
at position $\bf{r}$ in volume $\Omega_A$, associated with {\it
two electron current amplitudes} incident from contacts $1$ and
$2$. The explicit relation of the charge operator to local
wave functions is given in \cite{mb96} and a detailed
derivation is found in Ref. \cite{mb12}.

The density of states 
of the edge state in region $A$ and $B$ and the electrochemical 
capacitances are 

\begin{equation}
N_{\eta}=\sum_{\gamma} \, \,
{\cal N}^{\eta}_{\gamma\gamma}( {\bf r}),
\end{equation}
\begin{equation}
C_{\mu_{\eta}}=\frac{e^2N_{\eta}C_{\eta}}{C_{\eta}+e^2 N_{\eta}}.
\label{electro}
\end{equation}

The current correlation at equilibrium 
at a temperature $kT$ can be brought 
into the form
\be  
S_ {I_{\alpha}I_{\beta}}= 2 \omega^{2} 
C_{\mu_{\alpha}} C_{\mu_{\beta}} R_q^{\alpha \beta} kT \:,
\ee
where
\begin{eqnarray}
\label{rqgeneral} 
R_q^{\alpha \beta}= \frac{h}{2e^2} \frac{ \sum_{\gamma \delta}
{\rm Tr} \left[ {\cal N}_{\delta \gamma}^{(\alpha)} ({\cal
N}_{\delta \gamma}^{(\beta)})^{\dagger} \right]} {{\rm
Tr}\left[\sum_{\gamma} {\cal N}_{\gamma \gamma}^{(\alpha)} \right]
{\rm Tr} \left[\sum_{\gamma} {\cal N}_{\gamma \gamma}^{(\beta)}
\right]}.
\end{eqnarray}
For $\alpha = \beta = A$ these equations determine 
the auto-correlation and for $\alpha = A$
and $\beta = B$ the cross-correlation. 
In the zero temperature limit in the 
presence of an applied voltage $|eV|$
between contact $1$ and $2$ 
we can bring the current-current correlation 
into the form 
\be 
S_
{I_{\alpha}I_{\beta}}(\omega)= 2 \omega^{2} C_{\mu_{\alpha}} C_{\mu_{\beta}}
R_V^{\alpha \beta}  |e V| \:, 
\ee 
where
\begin{eqnarray}
\label{rvgeneral}
R_V^{\alpha \beta}= \frac{h}{2e^2} \frac{ {\rm Tr} \left[ {\cal
N}_{12}^{(\alpha)} ({\cal N}_{12}^{(\beta)})^{\dagger}
\right]+{\rm Tr} \left[ {\cal N}_{21}^{(\alpha)} ({\cal
N}_{21}^{(\beta)})^{\dagger} \right]} {{\rm Tr}
\left[\sum_{\gamma} {\cal N}_{\gamma \gamma}^{(\alpha)} \right]
 {\rm Tr} \left[\sum_{\gamma} {\cal N}_{\gamma \gamma}^{(\beta)} \right]}.
\end{eqnarray}
The electrochemical capacitances are positive and the 
sign of the current-correlations at the two gates 
is thus determined by $R_q^{\alpha \beta}$ at equilibrium 
and by $R_V^{\alpha \beta}$ in the zero temperature limit 
in the presence of transport. 
 
\begin{table}
\begin{center}
\renewcommand{\arraystretch}{2.0}
\begin{tabular}{|p{1.5cm}|p{1.5cm}|p{1.5cm}|p{1.5cm}|p{1.5cm}|}
\hline
   & ($i$) & ($ii$) & ($iii$) & ($iv$)\\ \hline
  $S^q_{I_A I_B}(\omega)$ & $>0$ & $=0$ & $\geq 0$  & $\geq 0$
\\
\hline $S^V_{I_A I_B}(\omega)$ & $\geq 0$ & $\leq 0$ & $=0$ & $=0$
\\
 \hline
\end{tabular}
\renewcommand{\arraystretch}{2.0}
\vspace{0.3cm} \caption[] {Sign of equilibrium ($S^q_{I_A
I_B}(\omega)$) and non- equilibrium ($S^V_{I_A I_B}(\omega)$)
current correlations between gates $A$ and $B$ for the four
positions of gate $B$ relative to gate $A$. \label{tb:1}}
\end{center}
\end{table}

For the geometry $I$ Ref. \cite{ammb} finds, 
\begin{equation}
R_q^{\alpha \beta}=(h/2e^2) \, \, \, \, \,{\rm and} \, \, \, \, \,
R_V^{\alpha \beta}=(h/e^2)TR, 
\label{eq:Rv1}
\end{equation}
independent of the choice of $\alpha$ and $\beta$. 
Here $T$ is the transmission probability through the quantum point 
contact and $R$ is the reflection probability. 
Thus at equilibrium the charge relaxation resistance is universal 
and given by $h/2e^2$. This results from the fact 
that a charge accumulated on the edge state near gate $A$ and $B$ 
can leave the sample only through contact $2$ where we have an interface 
resistance $h/2e^2$ . In the presence of transport, 
in the zero temperature limit considered here, the charge 
fluctuations reflect the shot noise and are proportional to $T(1-T)$. 
In geometry $I$ we find thus both at equilibrium and in the 
presence of shot noise a positive correlation. 

Consider next geometry $II$. Here gate $A$ and $B$ tests 
charge accumulation due to transmitted and reflected particles. 
These are mutually exclusive events and Ref. \cite{ammb} finds, 
\begin{eqnarray}
R_q^{AA}&=&R_q^{B B}=(h/2e^2), \\ R_q^{A B}&=&R_q^{B A}=0,
\\
R_V^{AA}&=&R_V^{B B}=-R_V^{A B}=-R_V^{B A}=(h/e^2)TR.
\label{eq:27}
\end{eqnarray}
The equilibrium correlations proportional to $R_q^{A B}$ are zero,
whereas the non-equilibrium correlations given by Eq.
(\ref{eq:27}) are negative. The results for the different geometries 
are summarized in Table $I$. 

The direct relation between charge fluctuations 
and the resistances $R_q$ and $R_v$ (see Eqs. (\ref{Four Parameters})
and Eqs. (\ref{rqgeneral},\ref{rvgeneral})) makes these quantities 
useful for many problems. Ref. {\cite{mb12,mbam,seelig} link 
these quantities to dephasing times in Coulomb coupled open conductors
and Ref.\cite{pilgram} demonstrates, that the dephasing time 
and relaxation time of a closed double quantum dot capacitively 
coupled to a mesoscopic conductor is governed by these resistances. 

\section{Cooper pair partition versus pair breaking noise} 

Hybrid-structures \cite{jehl1,kozh,jehl2,reulet,lefl}
consisting of a normal conductor 
and superconductor provide another system in which interactions 
play an important role in current-current 
correlations. 
At a normal-superconducting interface an electron (hole) 
is reflected as a hole (electron) if it is incident 
with an energy below the gap of the superconductor. 
This process, known as Andreev reflection, 
{\it correlates} excitations with different charge. 
Currents at the normal contacts of such a structure 
can be written as a sum an electron current $(e)$ and 
hole current $(h)$. Thus the correlation 
function $P_{\alpha\beta}$ can be similarly 
decomposed into four terms, 
\be 
P_{\alpha\beta} = P^{ee}_{\alpha\beta}
+ P^{hh}_{\alpha\beta} + P^{eh}_{\alpha\beta} + P^{he}_{\alpha\beta}
\ee
corresponding to correlations of currents of the same type
of quasi-particles and correlations between 
electron and hole currents. 
It can be shown that $P^{ee}$ and $P_{hh}$ are negative 
and $P^{eh}$ and $P^{he}$ are positive.  The sign of the correlation 
depends on the strength of the different 
contributions. Indeed Anantram and Datta \cite{anda}
showed that for a simple one-channel normal structure in which 
the normal part and the superconducting part form a loop 
penetrated by a flux $\Phi$, that the correlation measured 
at two normal contacts changes sign as a function of flux. 
Subsequent investigations 
based on a single channel Y-structure 
with a wave splitter which depends on a coupling 
parameter \cite{bia} found that the correlation 
changes sign and becomes positive as the coupling 
to the superconductor is decreased \cite{toma,lmb}. 
Investigation of a highly asymmetric 
geometry of an NS-structure in which one of the normal contacts 
is a tunneling tip found similarly restrictive 
conditions for positive correlations and moreover 
indicated that with increasing channel number of the normal 
conductor it is less and less likely to observe 
positive correlations \cite{gram99}. 

\begin{figure}[ht]
\vspace*{0.5cm}
\centerline{\epsfysize=5cm\epsfbox{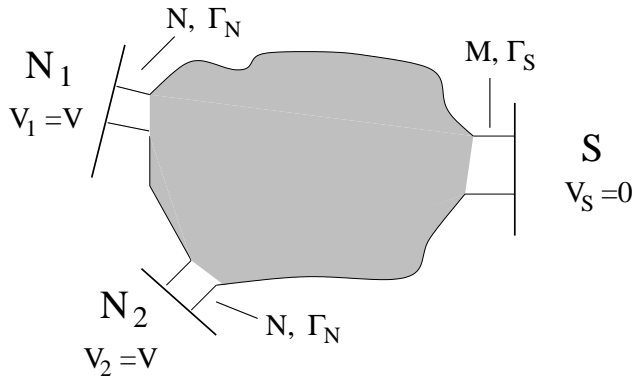}}
\vspace*{0.5cm}
\caption{\label{fig:sam1}
A chaotic cavity is connected
to two normal ($N_{1}$ and $N_{2}$) reservoirs and 
and to a superconductor $S$ via quantum point contacts.
After Samuelsson and B\protect\"uttiker \protect\cite{psmb1}.}
\end{figure}

These results pose the question of whether
positive correlations are indeed a feature of few channel ballistic 
systems only and could in fact not be seen in multi-channel 
systems which are typically also channel mixing. 
Indeed Nagaev and the author \cite{knmb} investigating diffusive 
normal structures, perfectly coupled to the superconductor, 
and neglecting the proximity effect, found that correlations 
are manifestly negative, as in purely normal structures. 
In view of the current interest in sources of 
entangled massive particles and the detection 
of entanglement, understanding the correlations
generated in hybrid structures is of 
particular interest \cite{sukh1,taddei,recher}.

To investigate the sign of current correlations 
in channel mixing hybrid structures for a wider range of conditions 
Samuelsson and the author \cite{psmb1} analyzed current correlations 
in a chaotic cavity using random matrix theory. 
The system is shown in Fig. \ref{fig:sam1}.
A chaotic cavity is coupled via quantum point contacts with $N_1 = N$ and 
$N_2 = N$ open channels at the normal contacts and with $M$ 
channels to the superconducting contact.

The result of the random matrix calculation 
is depicted in Fig. \ref{fig:sam2}. In the absence of 
the proximity effect (broken line in Fig. \ref{fig:sam2})
the ensemble averaged 
cross-correlation is negative over the entire range 
of the ratio $2N/M$. This situation 
is the analog for the chaotic cavity of the negative correlations 
found in diffusive conductors by Nagaev and the author \cite{knmb}. 
The result is dramatically different if the proximity  
effect plays a role (solid curve of Fig. (\ref{fig:sam2}). 
Now, at least in the limit where the cavity is much better 
coupled to the superconductor than to the normal reservoirs, 
the correlations are positive. Due to the multitude of processes 
contributing to these results a detailed microscopic explanation 
is difficult. In Ref. \cite{psmb1}
Samuelsson and the author present explanations 
for the limiting behavior ($N \ll M$ and $M\gg N$). 

A simple picture emerges if a barrier of 
strength $\Gamma_s$ is inserted 
into the contact between the cavity and a 
superconductor (see Fig. \ref{fig:sam1}). 
The case where there are tunnel barriers 
in all contacts has been investigated 
by Boerlin et al. \cite{boerlin}. 
Here we focus our attention to the case where 
the contacts to the normal reservoirs are perfect 
quantum point contacts and only the contact to 
the superconductor contains a barrier (as shown 
in the inset of Fig. {\ref{fig:sam3}). 
A simple result is obtained in the limit 
$2N/M \gg 1$. In this case 
injected quasi-particles scatter at most once 
from the superconductor-dot contact and 
the resulting scattering matrix simplifies considerably. 
The resulting correlation function is 
\be 
\frac{\langle P_{12}\rangle}{P_0}=\frac{M}{2N}R_{eh}(1-2R_{eh})
\label{notrssmallar}
\ee
where $R_{eh}=\Gamma_S^2/(2-\Gamma_S)^2$ is the Andreev reflection
probability of quasiparticles incident in the dot-superconductor
contact. There is a crossover from negative to positive correlations
that takes place
already for $R_{eh}=1/2$, i.e $\Gamma_S=2(\sqrt{2}-1)\approx 0.83$, in
agreement with the full numerics in Fig. \ref{fig:sam3}.
The fact that a "bad" contact reducing the Andreev reflection 
is favorable in generating positive correlations 
seems at first counter intuitive. Below we give 
a simple discussion to explain this result. 

\begin{figure}[ht]
\vspace*{0.5cm}
\centerline{\epsfysize=5cm\epsfbox{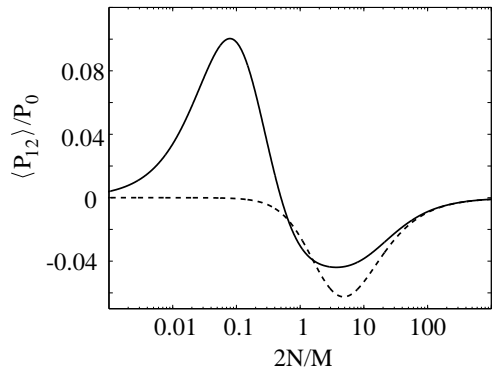}}
\vspace*{0.5cm}
\caption{\label{fig:sam2}
Current-current correlation of the chaotic dot junction coupled to the 
superconductor with $M$ open channels and with normal contacts 
of $N$ open channels as a function of $2N/M$. In the presence of 
the proximity effect (solid line) the correlation is positive 
for $2N < M$. 
Application of a magnetic flux of the order of a flux quantum 
suppresses the proximity effect and leads to negative correlations 
(broken line) independent of the ratio of $N$ and $M$. 
After P. Samuelsson and M. B\protect\"uttiker \protect\cite{psmb1}.} 
\end{figure}

Eq.(\ref{notrssmallar}) is the cross-correlation averaged 
over an ensemble of cavities. Since the proximity effect 
plays no role, it must be possible,  to derive this 
result from a purely semiclassical discussion. 
This statement holds of course not only for the particular 
geometry of interest here but of all the results obtained 
in the absence of the proximity effect. A semiclassical
theory for chaotic-dot superconductor systems 
is presented in Ref. \cite{psmb2} not only for the 
current-current correlations but also for the higher 
cumulants. Below we focus on the simple result described by 
Eq. (\ref{notrssmallar}). 

In the presence of the tunnel barrier at the superconductor-dot 
contact we can view the superconductor 
as an injector of Cooper pairs \cite{rsl}.
This picture differs from the Andreev-Bogoliubov-de Gennes
picture of correlated electron-hole processes. 
The (mathematical) transformation between these two pictures 
is of interest and will be discussed elsewhere. 
The argument presented below expands a suggestion 
by Schomerus \cite{schompriv}. 
We divide time into intervals such that the n-th time slot 
might contain a Cooper pair $\sigma_{n} = 1$ 
which has successfully penetrated through the barrier and entered 
the cavity or the n-th time slot is empty, $\sigma_{n} = 0$, 
if the Cooper pair 
has been reflected.  
Clearly, we have 
\be 
<\sigma_n>  \,= <\sigma^{2}_{n}> \,= R_{eh} 
\ee
where $R_{eh}$ is the Andreev reflection probability. 
Once the Cooper pair has entered the cavity two 
processes are possible: either the entire Cooper pair 
is transmitted into one of the normal contacts giving 
raise to {\it Cooper pair partition noise}
or the Cooper pair is split up and one electron 
leaves through contact $1$ and the other electron 
leaves through contact $2$. We refer to the contribution to 
the correlation function by this second process 
as {\it pair breaking noise}. To proceed 
we assume that each electron has a probability 
$T_1$ to enter contact $1$ and probability 
$T_2 = 1 -T_1$ to enter contact $2$. 
Thus for a symmetric junction $T_1 = T_2 = 1/2$
an incident Cooper pair contributes with probability $1/2$
to the partition noise and with probability $1/2$ 
to the pair breaking noise. 

\begin{figure}[ht]
\vspace*{0.5cm}
\centerline{\epsfysize=5cm\epsfbox{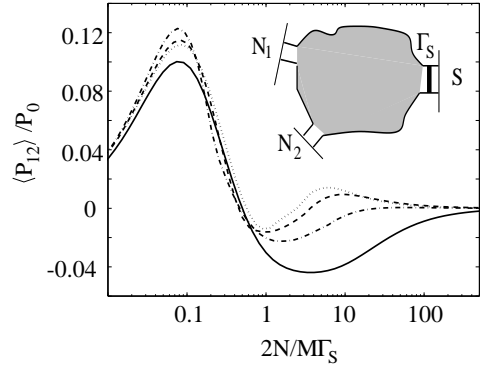}}
\vspace*{0.5cm}
\caption{\label{fig:sam3}
Current-current correlation of a chaotic cavity 
connected to the superconductor via a contact 
with conductance proportional to $\Gamma_s M$ 
and to two normal reservoirs with $N$ open channels 
as a function of $2N/M\Gamma_s$.
The contact transparencies are $\Gamma_s = 1$ (solid line)
$0.8$ (dashed dotted) $0.6$ (dashed line). For large 
$2N/M\Gamma_s$ the correlation crosses over from 
negative to positive as the transparency $\Gamma_s$ 
is reduced. 
After P. Samuelsson and M. B\protect\"uttiker \protect\cite{psmb1}.}
\end{figure}

We now want to write the correlation function 
in a way that permits us to separate these processes. 
The charge $Q_{1}$ transferred into contact $1$ over large number of 
time slots is 
\be 
Q_{1} = \sum_n \sigma_n (p_n + q_n)
\ee
and the charge $Q_{2}$ transferred into contact $2$ is 
\be 
Q_{2} = \sum_n \sigma_n (1 - p_n + 1- q_n) . 
\ee
Here $q_n$ and $p_n$ denote the two particles comprising the cooper pair. 
For pair partition we have $p_n = q_n =1$ or $p_n = q_n =0$ and 
for pair breaking we have $p_n = 1, q_n =0$ or $p_n = 0, q_n =1$. 

Next we consider the fluctuations of the transferred charge
$\Delta Q_{i} = Q_{i} - <Q_{i}>$. 
The average transmitted charge is $<Q_{i}> = R_{eh}$. 
For the correlation we find, 
\be
<\Delta Q_{1}\Delta Q_{2}> \,= \,<\Delta Q_{1}\Delta Q_{2}>_p + 
<\Delta Q_{1}\Delta Q_{2}>_e
\ee
where the index $p$ denotes the contribution 
of the pairs which are transmitted in their entierty 
into lead $1$ or $2$ 
and $e$ is the average only over the pairs 
which are broken up and an electron is emitted 
into each contact. 
For pair transmission we can distinguish events 
which emit a pair through the upper lead 1. 
In this case 
$\Delta Q_{1} = 2 \sum_n \sigma_n  - R_{eh}$
and 
$\Delta Q_{2} = - R_{eh}$
One quarter of all events are of this type. 
Similarly we can treat the case of pairs emitted through 
lead 2. Taking into account that 
$<\sum_n \sigma_n> = R_{eh}$ we find a pair partition 
noise 
\be
<\Delta Q_{1}\Delta Q_{2}>_p \,=\, - (1/2) R^{2}_{eh} 
\ee
Like the partition noise of single electrons in a normal 
conductor it is negative. 

Consider next the pair breaking events. 
For these events 
$\Delta Q_{1} = \sum_n \sigma_n - R_{eh}$
and 
$\Delta Q_{2} = \sum_m \sigma_m - R_{eh}$.
Onehalf of the time-slots with a Cooper pair are 
of this type. 
The correlation contains four terms,
$<\sum_n \sum_m \sigma_n\sigma_m> = R_{eh}$, 
$-R_{eh}<\sum_n \sigma_n> = - R_{eh} R_{eh}$, 
$-R_{eh}<\sum_n \sigma_n> = - R_{eh} R_{eh}$,
and a term $+R_{eh} R_{eh}$. 
Thus simultaneous emission of electrons 
gives rise to a noise 
\be
<\Delta Q_{1}\Delta Q_{2}>_e \,=\, (1/2) R_{eh}(1 - R_{eh}).
\ee
This contribution to the cross-correlation is positive.

Notice that the pair partition noise is negative and 
quadratic in $R_{eh}$. The pair breaking process gives a 
contribution which 
is linear in $R_{eh}$ for small $R_{eh}$ and thus wins in this 
limit. To achieve positive correlations it is thus 
favorably to have a small Andreev reflection probability. 
Only in this limit can the pair breaking processes 
overcome 
the negative partition noise of Cooper pairs
and give rise to positive correlations.  
The full counting statistics is discussed in Ref. \cite{psmb2}.

In hybrid superconducting normal structures 
there are thus several possibilities 
for a sign reversal of current-current fluctuations. 
For a cavity that is well coupled to a superconductor 
(see Fig. (\ref{fig:sam2}) application of a magnetic flux 
reverses the sign from positive to negative. 
As a function of temperature or applied voltage we can have 
a sign reversal both for the cavity that is well 
coupled to the superconductor (see Fig. \ref{fig:sam2})
as well as for the cavity that connects to the superconductor 
via a tunnel contact (see Fig. (\ref{fig:sam3}). For temperatures 
and voltages large compared to the superconducting gap 
the structure considered here behaves like a normal structure 
and exhibits negative correlations.


\section{Summary}

For non-interacting particles injected from thermal sources 
there is a simple connection 
between the sign of correlations 
and statistics. In contrast to photons, electrons 
are interacting entities, and we can expect the 
simple connection between statistics and the 
sign of current-current correlations to be broken, if interactions 
play a crucial role. 

The standard situation consists of a normal conductor embedded 
in a zero-frequency external impedance circuit such that the voltages 
at the contacts can be considered to be constant in time. 
Under this
condition the low frequency 
current-current cross-correlations measured at 
reservoirs are negative independent of the geometry, number of 
contacts and the bias applied to the conductor (as long as we 
do not depart to far from equilibrium).  
The negative correlations are a 
consequence of Fermi-statistics and the unitarity of the 
scattering matrix. Under these conditions the fluctuations 
in the potential play no role. We have shown that already the 
voltage fluctuations at a real voltage contact are sufficient 
to change the sign of correlations in certain special situations. 
Carriers injected by the voltage 
probe are correlated by the fluctuations of the potential 
of the voltage probe and can in the situation 
considered overcome the anti-bunching 
generated by Fermi statistics. 
We have also pointed out that displacement currents (screening currents) 
are positively correlated even at small frequencies. 
The electron-hole correlations generated in a normal conductor by 
a superconductor 
can similarly generate positive correlations in situations in which 
the pair partition noise is overcome by the pair breaking noise. 

The fact that interactions can have a dramatic effect on 
current-current correlations (change even their sign) 
clearly makes them a promising subject of further theoretical 
and experimental investigations.

\section*{Acknowlegements}

The work presented here is to a large extent
based on collaborations 
with Christophe Texier, Andrew Martin and Peter Samuelsson. 
I thank S. Pilgram, K. E. Nagaev and E. V. Sukhorukov for valuable 
comments on the manuscript. 
The work is supported by the Swiss National Science Foundation, 
the Swiss program for Materials with Novel Properties and 
the European Network of Phase Coherent Dynamics of Hybrid 
Nanostructures.

\end{document}